\documentstyle[amsmath,amssymb,graphicx]{article}

\def\be{\begin{eqnarray}}
\def\ee{\end{eqnarray}}
\def\nn{\nonumber}

\def\p{\partial}

\hoffset= - 1.0in         

\title{{\bf Nekrasov Functions and Exact Bohr-Sommerfeld Integrals
} \vspace{.2cm}}
\author{{\bf A.Mironov}\footnote{ {\small {\it
Lebedev Physics Institute} and {\it ITEP, Moscow, Russia}};
mironov@itep.ru; mironov@lpi.ru} \ and {\bf
A.Morozov}\thanks{{\small {\it ITEP, Moscow, Russia}};
morozov@itep.ru} \date{ }}

\begin{document}

\maketitle

\vspace{-5.0cm}

\begin{center}
\hfill FIAN/TD-21/09\\
\hfill ITEP/TH-51/09\\
\end{center}

\vspace{3.cm}

\begin{abstract}
In the case of $SU(2)$,
associated by the AGT relation to the $2d$ Liouville theory,
the Seiberg-Witten prepotential is constructed from
the Bohr-Sommerfeld periods of $1d$ sine-Gordon model.
If the same construction is literally applied to
monodromies of exact wave functions,
the prepotential turns into the one-parametric Nekrasov
prepotential ${\cal F}(a,\epsilon_1)$ with the other
epsilon parameter vanishing, $\epsilon_2=0$,
and $\epsilon_1$ playing the role of the Planck constant
in the sine-Gordon Shr\"odinger equation, $\hbar=\epsilon_1$.
This seems to be in accordance with the recent claim in
\cite{NeSha} and poses a problem of describing the
full Nekrasov function as a seemingly straightforward
double-parametric quantization of sine-Gordon model.
This also provides a new link between the Liouville and
sine-Gordon theories.
\end{abstract}

\section{Introduction}

The AGT conjecture \cite{AGT}, which is now explicitly checked and even
proved in various particular cases and limits
\cite{Wyl}-\cite{AGTlast},
provides a new prominent role for the Nekrasov functions \cite{Nek}.
Originally they appeared in description of regularized
integrals over moduli spaces of ADHM instantons \cite{insint},
but now it is getting clear that they provide a clever generalization
of hypergeometric series \cite{MMNek},
and thus can serve as a new class of special functions,
closely related to matrix model $\tau$-functions \cite{mamotau}.
This is a dramatic extension of the original role of
the Nekrasov functions and this means that they should be
thoroughly investigated within the general context of
group and integrability theory, without any references to particular
constructions like moduli spaces and graviphoton backgrounds.
There are several directions in which such study can be performed.
In the present paper we consider a possible way to embed the
Nekrasov functions into the context of Seiberg-Witten (SW) theory
\cite{SWfirst}-\cite{SWlast}, as suggested by N.Nekrasov and
S.Shatashvili in \cite{NeSha}.
We concentrate on the case of $SU(2)$ gauge group,
where Seiberg-Witten theory \cite{SWfirst} and its relation to quantum
mechanical integrable systems \cite{GKMMM} looks especially simple.
This allows one to formulate the claim of \cite{NeSha}
(as we understand it) in a very clear and transparent way,
what makes it understandable to non-experts in integrability theory.

According to \cite{GKMMM}, the SW prepotential \cite{SW2}
for the pure gauge $SU(2)$ ${\cal N}=2$ SUSY theory
is defined by the $1d$ sine-Gordon quantum model
\be
S = \int \left(\frac{1}{2}\dot\phi^2 - \Lambda^2\cos\phi\right) dt
\label{sG}
\ee
in the following way:
construct the Bohr-Sommerfeld periods
\be
\Pi^{(0)}(C) = \oint_C \sqrt{2(E-\Lambda^2\cos\phi})\ d\phi
\ee
for two complementary contours $C=A$ and $C=B$, encircling the
two turning points $\pm\phi_0$, $E=\gamma\cos\phi_0$.
Then, the SW prepotential ${\cal F}^{(0)}(a)$ is defined
from a pair of equations
\be
a = \Pi^{(0)}(A), \nn \\
\frac{\p{\cal F}^{(0)}(a)}{\p a} = \Pi^{(0)}(B)
\label{prepodef0}
\ee
after excluding $E$.
In the case of $SU(2)$ with a single modulus $a$, there is
no consistency condition for these equations to be resolvable,
however, it is slightly non-trivial that this construction is
directly generalized to higher-rank groups \cite{SWhighrank}.

The Bohr-Sommerfeld (BS) integrals are known to describe the quasiclassical
approximation $E^{(0)}$ to the eigenvalues $E$ of the Shr\"odinger
equation
\be
\left(-\frac{\hbar^2}{2}\frac{\p^2}{\p\phi^2}
+ \Lambda^2\cos\phi\right)\Psi(\phi)
= E\Psi(\phi)
\label{Shro}
\ee
by solving the equation
\be
\Pi^{(0)}(A) = 2\pi\hbar\left(n+\frac{1}{2}\right)
\ee
with respect to $E$.
The exact eigenvalues $E$ are defined from a similar equation
\cite{exBZ}
\be
\Pi(A) = 2\pi\hbar\left(n+\frac{1}{2}\right),
\ee
where the exact BS periods are
\be
\Pi(C) = \oint_C P(\phi) d\phi
\ee
and $P(\phi)$ is an exact solution to the Shr\"odinger equation
(\ref{Shro}),
\be
\boxed{
\Psi(\phi) = \exp\left(\frac{i}{\hbar} \int^\phi P(\phi) d\phi\right)
}
\ee
One can define the exact (quantized) prepotential ${\cal F}(a|\hbar)$
from the same system (\ref{prepodef0})
\be
\boxed{
\left\{
\begin{array}{l}
a = \Pi(A), \\ \\
\frac{\p{\cal F}(a|\hbar)}{\p a} = \Pi(B)
\end{array}
\right.
}
\label{prepodef}
\ee
with $\Pi^{(0)}$ substituted by the exact (quantized) periods $\Pi$.
Again, in the case of $SU(2)$ there is no problem of consistency
(resolvability) of this system.\footnote{
Note that, in variance with the quasiclassical $pdx$, the exact $Pdx$
is not a SW differential: its $a$-derivative is not
holomorphic on the spectral Riemann surface, moreover, there
is no smooth spectral surface anymore at all.}

The claim of \cite{NeSha} is that this ${\cal F}(a)$ is the
$\epsilon_2=0$ limit of the Nekrasov function,
\be
\boxed{
{\cal F}(a|\epsilon_1) = \lim_{\epsilon_2\rightarrow 0}
\Big\{\epsilon_1\epsilon_2\log Z(a,\epsilon_1,\epsilon_2)\Big\}
}
\label{F=Z}
\ee
so that $\epsilon_1$ plays the role of the Planck constant $\hbar$
in (\ref{Shro}).
The SW prepotential {\it per se} is \cite{Nek}
\be
{\cal F}^{(0)}(a) = {\cal F}(a|\epsilon_1=0)
\ee

{\bf Nota Bene:}
The deformation ${\cal F}^{(0)} \rightarrow {\cal F}$
is different from old Nekrasov's quantization \cite{Nek}
of the SW prepotential, in the direction
$\epsilon_2=-\epsilon_1$, which is AGT-related to
conformal models with integer central charge $c={\rm rank}$,
i.e. with $c=1$ in the $SU(2)$/Virasoro case.
The latter deformation,
associated with a background of self-dual graviphoton,
is a full topological partition function lifting
${\cal F}^{(0)}$ from zero to arbitrary genus \cite{DVAGT}.
It is a $\tau$-function which is
known to play a nice role in combinatorics
of symmetric groups \cite{comb,Hur}, but still lacks any nice
description in the simple terms of the sine-Gordon system (\ref{sG}).
Such a description is now found for the alternative deformation
in the direction of $\epsilon_1$, while $\epsilon_2=0$.
The two-parameter ($\epsilon_1,\epsilon_2$) deformation of
the SW prepotential, providing the full Nekrasov
function $Z(a,\epsilon_1,\epsilon_2)$ should be related
to a further, double-loop (elliptic or $p$-adic?), quantization
of the same sine-Gordon system.
One can see some evidence in support of this feeling in
the very recent paper \cite{index}.

\bigskip

Our goal in this letter is to demonstrate that
(\ref{prepodef}) and (\ref{F=Z}) are, indeed, correct
by calculating the first
orders of $\hbar$ expansion of the exact BS periods
and comparing them with the $\epsilon_1$-expansion of
the Nekrasov prepotential ${\cal F}(a|\epsilon_1)$
in the simplest case of $SU(2)$ theory.
We begin in s.\ref{prep}
with extracting ${\cal F}(a|\epsilon_1)$ from the Nekrasov functions.
We actually prefer to use the already established AGT relation
to describe ${\cal F}(a|\epsilon_1)$ in terms of the Shapovalov
matrix for the Virasoro algebra: this is a nice and clear
representation, which can be also used for other purposes.
Then in s.\ref{mono} we remind the old WKB construction
\cite{exBZ} of the exact (quantized) BS periods, which provides
exact eigenvalues of the Shr\"odinger equation, i.e. describe corrections
to the quasiclassical BS quantization rule, which we realize by action of
differential operators.
In s.\ref{BZsG} this construction is applied to the case
of sine-Gordon potential $V(\phi) = \Lambda^2\cos\phi$.
The simplest way to calculate the BS periods is with the
help of the Picard-Fucks equations \cite{IM,PF}, and corrections
are also obtained by action of peculiar differential operators.
Prepotential, defined from these corrected periods by the
SW rule, (\ref{prepodef}) does indeed coincide
with ${\cal F}(a|\epsilon_1)$ from s.\ref{prep},
provided one identifies $\hbar=\epsilon_1$.
The last section \ref{conc} contains a short summary and discussion.
Accurate proofs and numerous generalizations are left beyond
this letter to make presentation as clear as possible,
they are relatively straightforward and will be considered elsewhere.

\section{One parametric prepotential ${\cal F}(a|\epsilon_1)$
and Shapovalov matrix
\label{prep}}

The Nekrasov partition function for the $SU(2)$ pure gauge theory
possesses the nice group-theoretical description
\be
Z_{SU(2)}^{inst}(a,\epsilon_1,\epsilon_2) =
\sum_{n=0}^\infty \frac{\Lambda^{4n}}{(\epsilon_1\epsilon_2)^{2n}}
Q_\Delta^{-1}\big([1^n],[1^n]\big)
\label{ZQ}
\ee
where the Shapovalov matrix $Q$ is defined for a generic (non-degenerate)
Verma module of the Virasoro algebra with the central charge $c$
and the highest weight state $V_\Delta$.
Its elements are labeled by pairs of Young diagrams and given by
\be
Q_\Delta(Y,Y') = <L_{-Y}V_\Delta|L_{-Y'}V_\Delta>
= <V_\Delta | L_YL_{-Y'}V_\Delta>
\ee
$L_{-Y}$ is an ordered monomial made from negative Virasoro operators,
while the dimension and central charge are given by the AGT rule
\be
\Delta = \frac{1}{\epsilon_1\epsilon_2}\left(a^2-\frac{\epsilon^2}{4}\right),
\ \ \ \ \
c = 1 - \frac{6\epsilon^2}{\epsilon_1\epsilon_2},
\ \ \ \ \
\epsilon = \epsilon_1+\epsilon_2
\ee
$Q_\Delta(Y,Y')$ has a block form, it does not vanish only when the
two diagrams have the same size (number of boxes), $|Y|=|Y'|$.
Eq.(\ref{ZQ}) can be obtained as the large-mass limit of the
four-fundamentals AGT formula \cite{AGT,MMMagt}, see \cite{Glim,MMMlimf}
or, alternatively, as the large-$M$ limit of the adjoint AGT formula,
associated with the toric 1-point function \cite{AGTlast}
\be
Z_{adj}^{inst}(a,M,\epsilon_1,\epsilon_2) =
\sum_{Y,Y'} x^{|Y|} Q_\Delta^{-1}(Y,Y')
<L_{-Y}V_\Delta| L_{-Y'}V_\Delta(0)\ V_{\Delta_{ext}}(1)>
\ee
with $\Delta_{ext} = (M^2-\frac{\epsilon^2}{4})/(\epsilon_1\epsilon_2)$.
This formula (but not its large-$M$ limit) was recently
considered in \cite{Pog}.

In the limit of $\epsilon_2\rightarrow 0$ both the dimension $\Delta$
and the central charge $c$ tend to infinity, however, the singularities
are nicely combined and exponentiated, so that \cite{Nek}
\be
Z_{SU(2)}(a,\epsilon_1,\epsilon_2) =
\exp\left(\frac{{\cal F}(a,\epsilon_1,\epsilon_2)}{\epsilon_1\epsilon_2}
\right)
\ee
where ${\cal F}(a,\epsilon_1,\epsilon_2)$ remains finite when
$\epsilon_1\rightarrow 0$ or $\epsilon_2\rightarrow 0$.

Substituting explicit expressions for
\be
Q^{-1}_\Delta(1,1) = \frac{1}{2\Delta}\ \ - \ \ {\rm does\ not\ depend\ on}\ \ c,\nn\\
Q^{-1}_\Delta(11,11) = \frac{8\Delta+c}{4\Delta(16\Delta^2+2c\Delta-10\Delta+c)}  ,\nn \\
Q^{-1}_\Delta(111,111) = \frac{24\Delta^2+11c\Delta+c^2-26\Delta+8c}{24\Delta
(16\Delta^2+2c\Delta-10\Delta+c)(3\Delta^2+c\Delta-7\Delta+c+2)} ,\nn \\
\ldots
\ee
one easily obtains for the first terms of the $\Lambda$-expansion of
$
{\cal F}(a|\epsilon_1) \equiv {\cal F}(a,\epsilon_1,\epsilon_2=0)
= {\cal F}^{pert}(a|\epsilon_1) + {\cal F}^{inst}(a|\epsilon_1):
$
\be
{\cal F}^{inst}(a|\epsilon_1) =
\frac{\Lambda^4}{2\tilde\Delta}
+ \frac{\Lambda^8(10\tilde\Delta+6\epsilon_1^2)}
{16\tilde\Delta^3(8\tilde\Delta-6\epsilon_1^2)} + \ldots
= \left(\frac{\Lambda^4}{2a^2}
+ \frac{5\Lambda^8}{64 a^6} + \ldots\right)
+ \epsilon_1^2\left(\frac{\Lambda^4}{8a^4}
+ \frac{21\Lambda^8}{128a^8}
+ \ldots \right)
+ O(\epsilon_1^4)
\ee
where $\tilde\Delta$ denotes the rescaled $\Delta\to\epsilon_1\epsilon_2\Delta$.

It would also be interesting to describe ${\cal F}^{inst}(a|\epsilon_1)$
by taking the $\epsilon_2\rightarrow 0$ limit of coherent state
\cite{Glim,MMMlimf}, which provides an alternative description
of the pure gauge theory
\be
Z_{SU(2)}^{inst}(a,\epsilon_1,\epsilon_2) =
<\Lambda^2,\Delta|\Lambda^2,\Delta>
\ee
which satisfies
\be
L_0|\Lambda^2,\Delta>\ = \Delta|\Lambda^2,\Delta>,\nn \\
L_1|\Lambda^2,\Delta>\ = \Lambda^2|\Lambda^2,\Delta>,\nn \\
L_{k\geq 2}|\Lambda^2,\Delta>\ = 0
\ee

According to \cite{Nek,NeSha}, the perturbative contribution to
${\cal F}(a|\epsilon_1)$ is defined from its $a$-derivative,
\be
\frac{\p {\cal F}^{pert}}{\p a} =- 2\epsilon_1\log\left(\frac{\Gamma(1+z)}{\Gamma(1-z)}\right)
\ee
where $z = 2a/\epsilon_1$.
Making use of large-$z$ asymptotics of the $\Gamma$-function,
\be
\log\Gamma(z+1) = \log z + \log\Gamma(z) =
(z+1/2)\log z - z + \frac{1}{2}\log(2\pi) +
\sum_{m=1}^\infty \frac{B_{2m}}{2m(2m-1)z^{2m-1}}
\label{Gaexpan}
\ee
so that ($\ldots$ denotes here inessential terms)
\be
\log\Gamma(z+1) - \log\Gamma(1-z) =
2z\Big(\log z - 1  +
\sum_{m=1}^\infty \frac{B_{2m}}{2m(2m-1)z^{2m}}\Big) +...
\label{Gaexpansum}
\ee
one obtains
\be
-\frac{\p {\cal F}^{pert}}{\p a} =  8a\log\frac{\epsilon_1}{\Lambda}
+ 2\epsilon_1\log \Gamma\left(1+\frac{2a}{\epsilon_1}\right)
- 2\epsilon_1\log \Gamma\left(1-\frac{2a}{\epsilon_1}\right) = \nn \\ =
8a\left(\log\frac{2a}{\Lambda}-1\right) + {8a}\sum_{m=1}^\infty \frac{B_{2m}}{2m(2m-1)}
\left(\frac{\epsilon_1}{2a}\right)^{2m}+...
\ee
so that
\be
{\cal F}(a|\epsilon_1) = {\cal F}^{pert}(a|\epsilon_1)
+ {\cal F}^{inst}(a|\epsilon_1) = \nn \\
\boxed{
= -4a^2\log\frac{a}{\Lambda} -\frac{\epsilon_1^2}{6}\log a -
\left(\frac{\Lambda^4}{2a^2}
+ \frac{5\Lambda^8}{64 a^6} + \ldots\right)
- \epsilon_1^2\left(\frac{\Lambda^4}{8a^4}
+ \frac{21\Lambda^8}{128a^8}
+ \ldots \right)
+ O(\epsilon_1^4)
\label{Fexpan}
}
\ee
Note that only even powers of $\epsilon_1$ appear in this formula.

\bigskip

Our goal in this paper is to provide an alternative
description of this ${\cal F}(a|\epsilon_1)$
in terms of SW-like relation (\ref{prepodef}),
where $\Pi(C)$ are the exact BS periods
(monodromies of the exact wave function)
of the $0+1$ dimensional sine-Gordon model.

\section{Exact eigenvalues from quantized BS periods (monodromies)
\label{mono}}

Spectrum of the Shr\"odinger operator
$-\frac{\hbar^2}{2m}\frac{\p^2}{\p x^2} + V(x)$
is defined in the quasiclassical approximation by the BS
quantization rule
\be
\oint p_E(x) dx = \oint\sqrt{2m(E-V(x)}\ dx =
2\pi\hbar\left(n+\frac{1}{2}\right)
\ee
In fact, WKB theory allows one to calculate arbitrary corrections
to the quasiclassical approximation, up to any desired power
in $\hbar$.
Remarkably, exact $E$ is defined by the same quantization rule,
\be
\oint P_E(x)dx = 2\pi\hbar\left(n+\frac{1}{2}\right),
\label{quru}
\ee
only $p_E(x)$ should be substituted by $P_E(x)$, where
\be
\Psi_E(x) = \exp\left(\frac{i}{\hbar}\int^x P_Edx\right)
\ee
is the exact solution of the stationary Shr\"odinger equation,
\be
\left(-\frac{\hbar^2}{2m}\frac{\p^2}{\p x^2} + V(x)\right)
\Psi_E(x) = E\Psi_E(x)
\ee
In what follows we omit index $E$ from $p_E(x)$ and $P_E(x)$
to avoid further overloading formulas.

For thorough discussion of the quantization rule (\ref{quru})
see \cite{exBZ}, it can be justified by analysis of the
Stokes phenomenon and by study of the Airy function
asymptotic of $\Psi_E(x)$ in the vicinity of the turning points.
An advantage of this formula is that the $\hbar$ series for
$P(x)$ is constructed by a simple iteration:
substituting $P(x) = \sum_{k=0}^\infty \hbar^k P_k(x)$,
into $i\hbar P' = P^2-p^2$, one gets
\be
P_0(x) = p(x) = \sqrt{2m(E-V)}, \nn \\
P_1(x) = -i\frac{V'}{4(E-V)} = \frac{i}{4}\Big[\log(E-V)\Big]' =
\frac{i}{2}[\log P_0]', \nn \\
P_2(x) = \frac{1}{32}\frac{5V'^2 + 4V''(E-V)}{\sqrt{2m} (E-V)^{5/2}},\nn\\
P_3(x) = \frac{i}{64}\frac{4V'''(E-V)^2 + 18 V'V''(E-V) + 15V'^2}
{2m(E-V)^4} = i\left[\frac{P_2}{2P_0}\right]',\nn\\
\ldots
\label{ys}
\ee
In what follows we put $m=1$.

The energy levels are defined by the {\it exact} Bohr-Sommerfeld rule
\be
\oint P(x)dx = 2\pi\hbar (n+1/2)
\ee
i.e.
\be
\Pi = \frac{1}{\sqrt{2}}\oint P(x)dx =
\oint \sqrt{E-V} dx
-\frac{\hbar^2}{64} \oint \frac{V'^2dx}{(E-V)^{5/2}}
-\frac{\hbar^4}{8192} \oint \left(\frac{49V'^4}{(E-V)^{11/2}}
- \frac{16V'V'''}{(E-V)^{7/2}}\right)dx + \ldots
\ee
can be considered as a quantum deformation of the quasiclassical
periods
$\Pi^{(0)} = \oint \sqrt{E-V}dx$.
To simplify formulas, hereafter we divide periods by $\sqrt{2}$.

In this formula integration by parts is allowed and, therefore,
it looks simpler than (\ref{ys}).
Only $\hbar^{2k}$ corrections survive
($\hbar$ and $\hbar^3$ are indeed absent).
This fact will be important to match the absence of odd powers
of $\epsilon_1$ in (\ref{Fexpan}).

The $\hbar^2$-term can be alternatively represented as
\be
\Pi^{(2)} =
-\frac{\hbar^2}{64} \oint \frac{V'^2dx}{(E-V)^{5/2}} =
+\frac{\hbar^2}{96} \oint \frac{V''dx}{(E-V)^{3/2}} = \nn \\
= -\frac{\hbar^2}{96}\oint \frac{\gamma\cos \phi\ d\phi}
{(E-\gamma\cos\phi)^{3/2}} = -\frac{\gamma}{24}\p^2_{E\gamma}
\oint \sqrt{E-\gamma\cos\phi}\ d\phi =
-\frac{\hbar^2\gamma}{24}\p^2_{E\gamma} \Pi^{(0)}
\label{Pi2}
\ee
where in the second line we substituted the concrete potential of
the sine-Gordon model, $V(\phi) = \gamma\cos\phi$.

Similarly, integrating by parts one can rewrite the $\hbar^4$-term as
\be
\Pi^{(4)}= -\frac{\hbar^4}{3\cdot 2048}\oint\left(\frac{7V''^2}{(E-V)^{7/2}} +
\frac{2V''''}{(E-V)^{5/2}}\right)  = \nn \\
\ \stackrel{V=\gamma\cos\phi}{=}\
\frac{9\hbar^4}{128}\gamma\left(-\frac{2}{5}E\p_E + \gamma\p_\gamma\right)
\p^2_E\p_\gamma \oint \sqrt{E-\gamma\cos\phi}\ d\phi
\label{Pi4}
\ee

\section{Quantum corrections to BS periods in the sine-Gordon case
\label{BZsG}}

\subsection{Picard-Fucks equation \cite{IM,PF}}

The simplest way to evaluate the periods is to make use of the
Picard-Fucks equation \cite{IM,PF}
\be
\Big(\gamma (\p_E^2+\p_\gamma^2) + 2E\p^2_{E\gamma}\Big)\Pi^{(0)} = 0
\label{PFe}
\ee
This equation follows from the simple fact:
\be
\Big(\gamma (\p_E^2+\p_\gamma^2) + 2E\p^2_{E\gamma}\Big)
\sqrt{E-\gamma\cos\phi}\ d\phi =
\frac{2E\cos\phi - \gamma(1+\cos^2\phi)}{4(E-\gamma\cos\phi)^{3/2}}\
d\phi
= d\!\left(\frac{\sin\phi}{2\sqrt{E-\gamma\cos\phi}}\right)
\ee
We need to construct the two solutions of this equation with asymptotics
$\sqrt{2E} + O(\gamma)$ and $\sqrt{2E}\log(E/\gamma) + O(\gamma)$.
Since $E^{\frac{1}{2}+\varepsilon}
= \sqrt{E}\Big(1+\varepsilon\log E + O(\varepsilon^2)\Big)$
both periods can be obtained simultaneously,
by substituting into (\ref{PFe}) the formal series
\be
\Pi^{(0)}_\varepsilon
= \Pi^{(0)} + \varepsilon {\Pi^{(0)}}' + O(\varepsilon^2)
= \sqrt{2}E^{\frac{1}{2}+\varepsilon}\left(1+\sum_{n>0}
s_n\left(\frac{\gamma}{E}\right)^{2n}\right)
\ee
which provides a recursion relation
\be
s_{n+1} = \frac{\left(n+\frac{1}{4}-\frac{\varepsilon}{2}\right)
\left(n+\frac{1}{4}-\frac{\varepsilon}{2}\right)}
{(n+1)(n+1-\varepsilon)}\ s_n =
\left(\frac{n^2-\frac{1}{16}}{(n+1)^2} -
\frac{n+\frac{1}{16}}{(n+1)^3}\ \varepsilon
+ O(\varepsilon^2)\right)s_n
\ee
and
\be
\Pi^{(0)}_\varepsilon
= \sqrt{2E}\left( 1 - \frac{1}{16}\left(\frac{\gamma}{E}\right)^2 -
\frac{15}{2^{10}}\left(\frac{\gamma}{E}\right)^4 + \ldots \right) +\nn \\
+ \varepsilon\left\{\sqrt{2E}\log E
\left( 1 - \frac{1}{16}\left(\frac{\gamma}{E}\right)^2 -
\frac{15}{2^{10}}\left(\frac{\gamma}{E}\right)^4 + \ldots \right)
- \sqrt{2E}\left(\frac{1}{16}\left(\frac{\gamma}{E}\right)^2 +
\frac{13}{2^{11}}\left(\frac{\gamma}{E}\right)^4 + \ldots\right)\right\}
+ O(\epsilon^2)
\ee
According to Seiberg-Witten theory, we identify
$\Pi^{(0)} = a$, ${\Pi^{(0)}}' = 1/4\p F_{SW}(a)/\p a$.
It follows that
\be
\sqrt{2E} = a\left(1 + \frac{1}{4}\left(\frac{\gamma}{a^2}\right)^2
+ \frac{3}{64}\left(\frac{\gamma}{a^2}\right)^4 + \ldots\right)
\label{Evsa}
\ee
and substituting this into ${\Pi^{(0)}}'$, we get:
\be
-{1\over 4}  \frac{\p F_{SW}(a)}{\p a} = {\Pi^{(0)}}' =
2a\log a + \frac{1}{4}\, a\!\left(\left(\frac{\gamma}{a^2}\right)^2 +
\frac{15}{32}\left(\frac{\gamma}{a^2}\right)^4 + \ldots\right)
\label{FSWa}
\ee
i.e.
\be
F_{SW}(a) = -4a^2(\log a + const) - \frac{\gamma^2}{2a^2}
-\frac{5\gamma^4}{64a^6} + \ldots
\label{FSW}
\ee
This is a well-known formula in Seiberg-Witten theory.
Since $\gamma=\Lambda^2$,
one sees that it is in accordance with the $\epsilon_1$-independent
term in formula (\ref{Fexpan}) obtained entirely within
conformal field theory, from the 1-point toric conformal block.

\subsection{Prepotential in the order $\hbar^2$}

According to (\ref{Pi2}),
acting by the operator $\left(1 - \frac{\hbar^2\gamma}{24}\p^2_{E\gamma}\right)$
on $\Pi^{(0)}$, one obtains
\be
\Pi^{(0)}_\varepsilon + \Pi^{(2)}_\varepsilon =
\left(1 - \frac{\hbar^2\gamma}{24}\p^2_{E\gamma}\right)\left[
\sqrt{2E}\left( 1 - \frac{1}{16}\left(\frac{\gamma}{E}\right)^2 -
\frac{15}{2^{10}}\left(\frac{\gamma}{E}\right)^4 + \ldots \right) +
\right.\nn \\ \left.
+ \varepsilon\left\{\sqrt{2E}\log \frac{E}{\gamma}
\left( 1 - \frac{1}{16}\left(\frac{\gamma}{E}\right)^2 -
\frac{15}{2^{10}}\left(\frac{\gamma}{E}\right)^4 + \ldots \right)
- \sqrt{2E}\left(\frac{1}{16}\left(\frac{\gamma}{E}\right)^2 +
\frac{13}{2^{11}}\left(\frac{\gamma}{E}\right)^4 + \ldots\right)\right\}
+ O(\epsilon^2)\right] =\nn\\
= \sqrt{2E}\left( 1 - \frac{1}{16}\left(\frac{\gamma}{E}\right)^2
\left(1 + \frac{\hbar^2}{8E}\right)
-\frac{15}{2^{10}}\left(\frac{\gamma}{E}\right)^4
\left(1 + \frac{7\hbar^2}{12E}\right) + \ldots\right) + \nn \\
+ \varepsilon\left\{\sqrt{2E}\log \frac{E}{\gamma}
\left( 1 - \frac{1}{16}\left(\frac{\gamma}{E}\right)^2
\left(1 + \frac{\hbar^2}{8E}\right)
-\frac{15}{2^{10}}\left(\frac{\gamma}{E}\right)^4
\left(1 + \frac{7\hbar^2}{12E}\right)
+ \ldots \right) - \right.\nn \\ \left.
- \sqrt{2E}\left(\frac{1}{16}\left(\frac{\gamma}{E}\right)^2
\left(1 + \frac{\hbar^2}{8E}\right)
+\frac{13}{2^{11}}\left(\frac{\gamma}{E}\right)^4
\left(1 + \frac{7\hbar^2}{12E}\right) + \ldots\right)
+\right. \nn \\ \left.
+ \frac{\hbar^2\sqrt{2E}}{24E}
\left(\frac{2}{16}\left(\frac{\gamma}{E}\right)^2 +
\frac{15}{2^{8}}\left(\frac{\gamma}{E}\right)^4
+\ldots\right)
+ \frac{\hbar^2\sqrt{2E}}{48E}
\left(1+\frac{3}{16}\left(\frac{\gamma}{E}\right)^2 +
\frac{105}{2^{10}}\left(\frac{\gamma}{E}\right)^4
+\ldots\right)
\right\} + O(\epsilon^2)
\ee
The two terms in the last line come from differentiation of
$\log E$ and $\log\gamma$ respectively.

Expressing $E$ through $a$ by solving the equation $\Pi = a$
and substituting the result into $\Pi'$ one obtains instead
of (\ref{Evsa}) and (\ref{FSWa})
\be
\sqrt{2E} = a \left\{ 1+\frac{1}{4}\left(\frac{\gamma}{a^2}\right)^2
\left(1+\frac{\hbar^2}{4a^2}\right) +
\frac{3}{64}\left(\frac{\gamma}{a^2}\right)^4 +
\frac{19\hbar^2}{128a^2}\left(\frac{\gamma}{a^2}\right)^4
+ O(\gamma^6,\hbar^4)\right)
\ee
and
\be
-{1\over 4}  \frac{\p F(a,\hbar)}{\p a} = {\Pi^{(0)}}' =
2a\log a + \frac{\hbar^2}{12a} +
\frac{1}{4}\, a\!\left\{\left(\frac{\gamma}{a^2}\right)^2
\left(1 + \frac{\hbar^2}{2a^2}\right)
+\frac{15}{32}\left(\frac{\gamma}{a^2}\right)^4
+\frac{21\hbar^2}{16}\left(\frac{\gamma}{a^2}\right)^4
+ \ldots\right\}
\label{FSWa1}
\ee
and finally, instead of (\ref{FSW}),
\be
\boxed{
F(a,\hbar) = -4a^2(\log a + const) - \frac{\hbar^2}{3}\log a
- \frac{\gamma^2}{2a^2} - \frac{5\gamma^4}{64a^6}
- \frac{\hbar^2\gamma^2}{8a^4} - \frac{21\hbar^2\gamma^4}{128a^8}
+ O(\gamma^6,\hbar^4)
}
\label{FSW1}
\ee
This reproduces  (\ref{Fexpan}),
provided one identifies $\gamma = \Lambda^2$ and $\hbar = \epsilon_1$.

\subsection{Prepotential in the order $\hbar^4$}

In this order we perform calculation
only for the $\gamma$-independent terms in
${\cal F}(a|\epsilon_1)$. Such contributions come from the
action on $\sqrt{2E}\log\frac{2E}{\gamma}$ in $\Pi^{(0)}$:
\be
\left\{1 - \frac{\hbar^2}{24}\gamma\frac{\p^2}{\p E\p \gamma} +
\frac{\hbar^4}{9\cdot 128}\gamma\left(-\frac{2}{5}E\p_E
+ \gamma\p_\gamma\right)
\frac{\p^3}{\p E^2\p \gamma} + \ldots\right\}
\sqrt{2E}\log\frac{2E}{\gamma} = \nn \\
= \sqrt{2E}\left(\log\frac{2E}{\gamma}
+ \frac{\hbar^2}{48E} - \frac{\hbar^4}{32\cdot360E^2}
+ \ldots \right) =
\boxed{
2 \sqrt{2E}\left(\log\frac{2E}{\gamma}
+ \frac{B_2}{2}\left(\frac{\hbar^2}{8E}\right)
+ \frac{B_4}{12}\left(\frac{\hbar^4}{8E}\right)^2
+ O(\gamma^2,\hbar^6) \right)
}
\ee
with $B_2=1/6$, $B_4=-1/30$.
This is again in agreement with (\ref{Fexpan}) if
$\frac{\hbar^2}{8E} = \frac{\epsilon_1^2}{(2a)^2} +O(\gamma)$.
Since $2E=a^2 + O(\gamma)$ this implies that $\hbar=\epsilon_1$.

This completes our simple test of the claim (\ref{F=Z}).

\section{Conclusion
\label{conc}}

In this letter we explicitly demonstrated that the deformation
from Seiberg-Witten prepotential to the Nekrasov function
continues to be described by the integrable system approach
suggested in \cite{GKMMM},
at least, for a 1-parametric
deformation to arbitrary $\epsilon_1$, with $\epsilon_2=0$.
The deformed prepotential ${\cal F}(a|\epsilon_1)$ is given by
exactly the same SW rule (\ref{prepodef}), only the SW differential
$pdx$ is deformed into the exact "quantum" differential $Pdx$.
The main open question is what should be done with the same
description when both $\epsilon_1$ and $\epsilon_2$ are
non-zero.
This generalization should be similar to the next, "elliptic"
deformations of quantum groups, which are themselves deformations
of the ordinary universal enveloping algebras.
As usual \cite{GLM}, one could expect that such deformations
will be also related to double loop algebras and to $p$-adic
analysis.

\bigskip

There is also a number of technical questions at the level
of ${\cal F}(a|\epsilon_1)$.

First of all, even in the case
of $SU(2)$ we presented only the lowest terms of $\epsilon_1$
expansion, one can look at generic terms and find a general
proof of the statement.

Second, it can be easily generalized
to other $SU(2)$ examples, that is, from the sine-Gordon to
Calogero-Ruijenaars and magnetic systems, which are integrable
system counterparts of various gauge theories under the
GKMMM-DW correspondence of \cite{GKMMM} and \cite{DW}.

Third, one should sum up the $\hbar$ series for $Pdx$, at least,
conceptually, e.g. interpret deformed differential $Pdx$
as a SW differential on a deformed (quantized) spectral
curve.

Forth, generalization to higher rank groups requires a more
detailed analysis. A piece of such analysis is presented in
\cite{NeSha} in terms of advanced integrability theory
{\it a la} \cite{intth},
but it is desirable to convert it into a much simpler
form, close to the one in the present paper.\footnote{
Another problem with the presentation of \cite{NeSha} is that
only the second of the two equations
in (\ref{prepodef}) is actually considered there
(in somewhat different terms).
This is enough to compare with the Nekrasov functions
which already have the proper $a$ as their argument,
but not enough for the SW construction,
where $a$ still needs to be defined. }
The simplest option is to construct an $\hbar$-deformed
SW differential and define ${\cal F}(a|\epsilon_1)$ from
the system (\ref{prepodef}) with $2\times{\rm rank}$
different periods.
For example, in the case of $SU(N)$ gauge theory,
associated {\it a la} \cite{GKMMM} with the $N$-body
affine Toda model,
the SW prepotential ${\cal F}(\vec a)={\cal F}(a_1,\ldots,a_{N-1})$
is defined through (\ref{prepodef}) by the $2N-2$ periods
of the SW differential $pdx$, where \cite{SWhighrank}
\be
p^N - \sum_{k=0}^{N-2} E_kp^k = 2\Lambda^2\cos x
\ee
while ${\cal F}(\vec a|\epsilon_1)$ can be similarly defined
through the same (\ref{prepodef}) by the $2N-2$ periods of
the deformed differential $Pdx$, which appears in solution
$\Psi = \exp\left(\frac{i}{\hbar}\int^x Pdx\right)$
to the Baxter equation
\be
\left\{\big(-i\hbar\p_x\big)^N
- \sum_{k=0}^{N-2} E_k\big(-i\hbar\p_x\big)^k - 2\Lambda^2\cos x
\right\} \Psi(x) = 0
\ee
with $\hbar=\epsilon_1$. We remind \cite{intth} that
the Fourier transform of this equation arises after separation of variables from
the Shr\"odinger equation for $N$-body Toda theory.
In the SW case, consistency of (\ref{prepodef}) for $N>2$
is guaranteed by the holomorphicity of the differentials
$\frac{\p (pdx)}{\p E_k}$ and the symmetry $T_{jk} = T_{kj}$
of the period matrix. In the deformed case, the idea can be
that integration contours encircle all the singularities of $ydx$.
The deformed SW differential $Pdx$ will be obtained
from the original one $pdx$ by an action of operators like
(\ref{Pi2}) and (\ref{Pi4}), whose explicit form remains
to be found.

Fifth, since the AGT relation expresses the Nekrasov functions in terms
of conformal blocks in $2d$ Liouville-Toda theories,
the whole construction provides a new relation between
open and periodic Toda systems, in particular, between Liouville
and sine-Gordon models. It is well known that the free field
formulation of Liouville theory {\it a la} \cite{DF}
requires its lifting to the sine-Gordon model, it would be nice to
find an explicit connection between that construction and the one
in the present paper, see also \cite{MMNek} for a related
set of questions.
Note that the BS description of the prepotential ${\cal F}(a|\epsilon_1)$
unavoidably contains its perturbative part and thus is sensitive
to the choice of conformal model, not only to the chiral algebra.

\section*{Acknowledgements}

We are grateful to S.Kharchev for valuable discussions.

The work was partly supported by Russian Federal Nuclear Energy
Agency and by RFBR grants 07-02-00878 (A.Mir.),
and 07-02-00645 (A.Mor.).
The work was also partly supported
by joint grants 09-02-90493-Ukr,
09-02-93105-CNRSL, 09-01-92440-CE, 09-02-91005-ANF and by Russian President's Grant
of Support for the Scientific Schools NSh-3035.2008.2.

\end{document}